\documentclass[sigconf]{acmart}

\usepackage{booktabs} 
\usepackage{url}
\usepackage{graphicx}
\usepackage{caption}
\usepackage{amsmath}
\usepackage{threeparttable}
\usepackage{amssymb}
\DeclareCaptionType{InfoBox}
\usepackage{tikz}

\usetikzlibrary{fadings}

\tikzset{fading text/.style={}}

\definecolor{blue1}{RGB}{139,157,195}

\definecolor{blue2}{RGB}{247,247,247}

\newcommand\fadingtext[2][]{%
	\begin{tikzfadingfrompicture}[name = fading letter]
		\node[color=black, inner xsep = 0pt, outer xsep = 0pt] {#2};
	\end{tikzfadingfrompicture}%
	\begin{tikzpicture}[baseline = (textnode.base)]	
	\node[inner sep = 0pt, outer sep = 0pt, #1] (textnode) {#2}; 	
	\shade[path fading = fading letter, fading text, #1,fit fading = false]
	(textnode.south west) rectangle (textnode.north east);%
	\end{tikzpicture}%
}

\begin{document}
\title{Attention based Sentence Extraction from Scientific Articles using Pseudo-Labeled data}

\author{Parth Mehta}
\affiliation{%
  \institution{DA-IICT}
  \streetaddress{Gandhinagar}
}
\email{parth\_me@daiict.ac.in}

\author{Gaurav Arora}
\affiliation{%
	\institution{DA-IICT}
	\streetaddress{Gandhinagar}
}
\email{arora\_gaurav@daiict.ac.in}

\author{Prasenjit Majumder}
\affiliation{%
	\institution{DA-IICT}
	\streetaddress{Gandhinagar}
}
\email{p\_majumder@daiict.ac.in}

\begin{abstract}
In this work, we present a weakly supervised sentence extraction technique for identifying important sentences in scientific papers that are worthy of inclusion in the abstract. We propose a new attention based deep learning architecture that jointly learns to identify important content, as well as the cue phrases that are indicative of summary worthy sentences. We propose a new context embedding technique for determining the focus of a given paper using topic models and use it jointly with an LSTM based sequence encoder to learn attention weights across the sentence words. We use a collection of articles publicly available through ACL anthology for our experiments. Our system achieves a performance that is better, in terms of several ROUGE metrics, as compared to several state of art extractive techniques. It also generates more coherent summaries and preserves the overall structure of the document.
\end{abstract}

%
%


\maketitle

\section{Introduction}

Automatic text summarization has been a major focus area of researchers for a few decades now. However, due to small amount of training data and even fewer number of gold standard summaries, experiments in automatic summarization have largely been dependent on the newswire corpora created under the \emph{document understanding conference (DUC)} and \emph{text analysis conference (TAC)}. With the increasing use of deep learning techniques for various NLP tasks, large volumes of training data are more important than ever before, but few of them are publicly available\cite{rush2015neural},\cite{filippova2015sentence}. In this work we draw attention to one such publicly available corpus of scientific articles published in ACL anthology. This is perhaps one of the largest corpora of documents with corresponding manually written abstract. The only other corpora of similar size come from newswire data. But the summaries, in that case, are largely limited to highlights or article headings as opposed to an actual abstractive summary. All the previous experiments using scientific articles were limited to not more than a few hundred articles\cite{abu2011coherent}. 

In this work, we propose a novel sentence extraction technique as a first step towards automatically generating the abstracts of scientific papers. We demonstrate that a very efficient sentence extractor can be created using this data, with weakly supervised training and without any manual labelling. The main contributions are twofold, firstly we propose a simple context encoder which can capture the overall theme of the document and generate a context embedding. Second, we propose an attention model that uses sequence encoder based sentence embeddings along with this context embedding to assign importance to different words in a given sentence. This module jointly learns to capture the informative content as well as the cue phrase information that make a sentence summary worthy. As we show in the results, our model is able to identify and leverage cue phrases like '' and '' to decide the \emph{summary worthiness} of the sentence. Using this information, we are able to maintain the overall structure of document when creating the abstract, which is not possible using the existing extractive techniques. Contrary to most of the existing techniques, our approach is not dependent on manually tagged data or any linguistic resources.
 
One of the most notable attempt at generating abstracts from scientific articles is by \cite{teufel2002summarizing}. They solve this problem by leveraging rhetorical status of sentences to generate the abstract. The idea is to select sentences in such a manner that the abstract highlights new contribution of the paper and also relates it to the existing work. The authors identify \emph{Rhetorical zones} to which a sentence can belong like: the aim of the experiment, statements that describe the structure of article, comparison with other works, etc. Such features are then used to decide the importance of the sentence as well as to maintain the overall structure of the final extract. The work by \cite{mei2008generating} focuses on generating impact based summaries for scientific articles. Sentences are ranked based on the impact they have produced on other works in the same or related domains. \emph{Document sentences} that best match the content of the \emph{citing sentence} are identified using language models. They used a dataset of around 1300 papers published by ACM SIGIR. The work described in \cite{abu2011coherent} clusters articles based on their citation graph and then use lexank to rank sentences within each cluster. The work proposed in \cite{hirohata2008identifying} focuses on identifying the sections of the original paper to which a given abstract sentence is related. Our model implicitly tries to learn similar information. In the results, we show that the attention model learns to identify phrases which are indicative of the section information and such sentences are usually selected in the summary. Another related work to the proposed approach is by \cite{cheng2016neural}. It focuses on generating headlines of news articles using sentence and document encoders. The authors use sentence and word level labelling to identify important phrases in the document and then generate an extract based on that technique. 

\section{Proposed Architecture}

Our proposed model consists of four main blocks: An LSTM based sentence encoder, topic modelling based context encoder, attention module and a binary classifier. Overall the aim is to determine the probability $p(y|s,d)$ where $p(y)$ is the probability that sentence $s$ in a document $d$ is \emph{summary worthy}. We represent $s$ as an embedding vector of fixed dimensions, using a sentence encoder. Next, we represent each document by the topic extracted using LDA, and use those topics to create a context embedding. The attention model then uses the sentence and context embeddings to learn to assign weights to different parts of the sentence. Finally, the classifier uses the output of attention module and the original context embeddings to decide whether or not a sentence is \emph{summary worthy}. Below we describe the individual blocks.

\subsection{Sentence Encoder} 
Each sentence $S$ is represented as a sequence of $N$ vectors $[\boldsymbol{x}_1$,...$\boldsymbol{x}_N]$ where $\boldsymbol{x}_i$ is the $i^{th}$ word represented by its word embedding vector. The initial word embeddings were created by training a word2vec\cite{mikolov2013efficient} model on the entire ACL corpus and were updated during the training. The word embedding matrix $\boldsymbol{E}$ is of size $V\times D$, where $V$ is the vocabulary size and $D$ is the word embedding size.
Next, we use an LSTM based sequence encoder with a hidden size of $U$ for creating the sentence embeddings using these word embeddings. LSTM based sentence encoders are now considered a standard technique for creating sentence embeddings. We limit the maximum possible length of a sequence to $L$.

\subsection{Context Encoder} Even for humans, knowledge about the overall scope of an article is pivotal when selecting important information that has to be included in the abstract. There have been attempts to generate a document encoding, by using an additional LSTM based sequence encoder that takes input a sequence of sentence embeddings created by the sentence encoder defined above\cite{cheng2016neural} and gives a single vector or the \emph{document embedding}. However, such an approach requires a large amount of training data, of the order hundreds of thousands of article, and takes much longer to train. As an alternative, we propose a simpler approach, that efficiently captures the overall scope of the document and can be efficiently trained using a few thousand documents. It is noteworthy that here our aim is not to capture the document structure explicitly but to capture the overall theme of the document.

Our context encoder follows a two step approach. In the first step, we encode each article in terms of representative concepts present in them. We extracted 500 abstract topics from the overall corpus using \emph{Latent Dirichlet Allocation} based topic modelling. Topic vectors for each document can be represented as a matrix  $T \in \mathbb{R}^{M\times M}$, $\boldsymbol{T}$ = $[t_1,...,t_M]$, where $t_i$ is the one-hot encoded vector of size $1 \times M$ for topic $i$, and $M$ is the pre-decided number of topics. We separately initialized a topic embedding matrix $\boldsymbol{F} \in \mathbb{R}^{M\times C}$, where $M$ is the total number of topics and  $C$ is the context embedding size. We randomly initialize $\boldsymbol{F}$ and it is jointly updated with the overall model. $J \in \mathbb{R}^{C\times M}$ represents the topic embeddings. We then perform a weighted average of the topic embeddings using their probabilities($\boldsymbol{p_i}$). This additional step helps in reducing the sparsity of LDA representation as well as to leverage latent similarity between different topics, and at the same time assigning an importance score to each of the topics. $c \in \mathbb{R}^{C\times 1}$ represents the final weighted context embeddings.
\vspace{-1mm}
\begin{eqnarray}
J =  F^\intercal T \\
c = \sum\limits_{i} p_i J_i
\label{eq3}
\end{eqnarray}
\vspace{-3mm}
\subsection{Attention module} This module plays a key role in the overall architecture. In past few years, attention networks have become a popular choice for several NLP tasks. Several approaches have been proposed for using attention for document summarization\cite{cheng2016neural},\cite{rush2015neural}. We propose a simple attention architecture that takes into account the document context and sentence embedding for generating attention weights over the sentence words. We argue that besides identifying informative content in the document, such an attention model would help in automatically identifying words or phrases, which can act as a cue for deciding whether or not that sentence is summary worthy. The attention weights($[w_1,...,w_L]$) are computed as shown in equation 3, where $Z \in \mathbb{R}^{(S+C)\times L}$ and $w \in \mathbb{R}^{L\times1}$. The attention module learns weights $\boldsymbol{w}$ as a function of the sentence embedding(local context) as well as the context embedding (global context). $L$ is the maximum allowed length of an input sentence. Sentences shorter than this are padded to make them of the same length. $Y \in \mathbb{R}^{L\times S}$ denotes the intermediate steps of LSTM output at each of the L timestamps. $Y = [y_i,...y_L]$ where $\boldsymbol{y_i}$ represents intermediate output at a particular time stamp $i$. $s = y_L$.
\vspace{-1mm}
\begin{eqnarray}
\boldsymbol{w} = Z(s,c)\\
\label{eq2}
\boldsymbol{a} = w^\intercal Y
\end{eqnarray}
\vspace{-5mm}
\subsection{Classifier} The classifier consists of two layered feed forward network. We used a hidden layer with weights $\boldsymbol{H} \in \mathbb{R}^{(A+C) \times Q}$ followed by a output layer $\boldsymbol{O} \in \mathbb{R}^{Q \times 1}$ and a sigmoid activation function ($\sigma$). 

\begin{eqnarray}
h = \boldsymbol{H}[a,c]\\
o = \sigma(\boldsymbol{O}h)
\end{eqnarray}

The entire architecture is shown in the Figure \ref{fig1} below.
\begin{figure}
\includegraphics[scale=0.25]{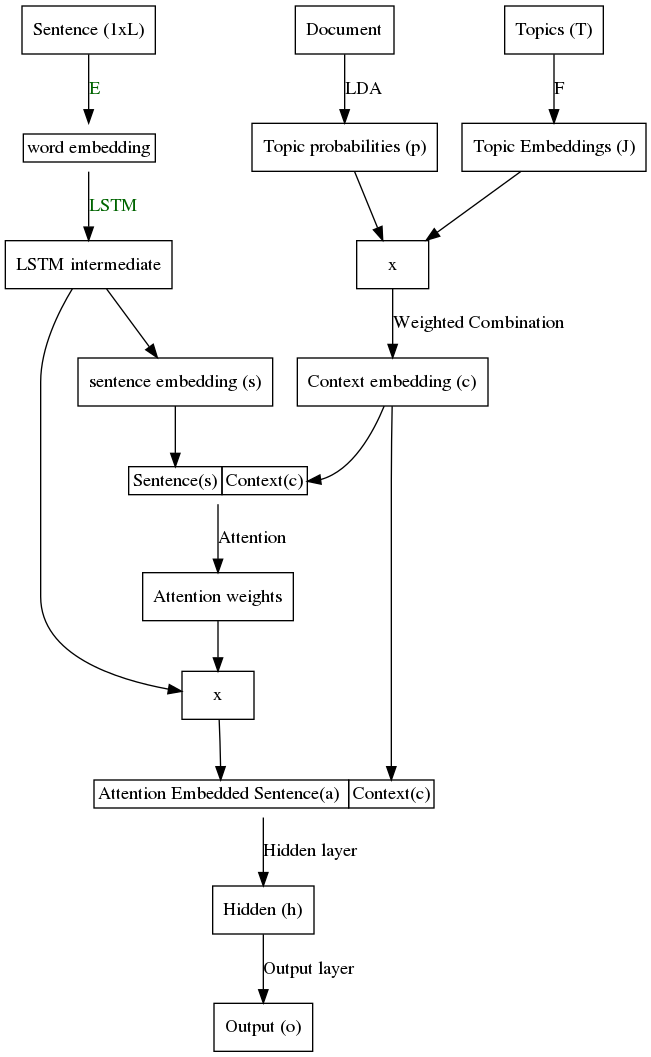}
\caption{Attention based sentence selector}	
\label{fig1}
\end{figure}

\vspace{-2mm}
\section{Experimental Setup}

We use a subset of the ACL Anthology corpus which is a collection of scientific articles broadly from computational linguistics and related domains. These articles are openly available in the ACL anthology website\footnote{\url{http://aclweb.org/anthology/}} in pdf formats. We used the publicly available Science Parse library\footnote{\url{https://github.com/allenai/science-parse}} for extracting section wise information from the pdf documents. Only the articles published in or after the year 2000 were included. Further, the articles that were not parsed properly were discarded. Finally, 27,801 articles were used in this experiment, which were divided into train (23000), validation(2000) and test sets(2801). The ids of these articles can be found here\footnote{\url{url unavailable to maintain anonymity}}. We did not perform any preprocessing or manual labelling. 

For each sentence in the document, we assign a pseudo-label of 1(\emph{important}) or 0(\emph{not important}), based on their cosine similarity with the sentences in abstract. For each sentence in the abstract, we select the best matching sentence from the document if the cosine similarity is above 0.75 and assign it a label 1. All other sentences are assigned a label 0. Compared to a summary of a newswire cluster, abstracts of scientific articles are much more precise with a higher compression ratio. The average size of ACL articles is 200 sentences or about 3600 words, while the average summary size is 125 words. This results in a heavily skewed training data, with more than 95\% sentences being labelled as \emph{not important}. To mitigate this bias, we filter out sentences with tf-idf scores lower than 0.05. Further, we randomly sample the \emph{not important} sentences to bring down the positive-negative ratio to 1:4. We then use a weighted loss function, explained below, to assign a higher loss to false negatives as compared to false positives. Next we train the attention based model described in the previous section with the sentences as inputs and the pseudo labels as the target. The implementation details and choice of parameters are described below.

\subsection{Implementation Details} 

We used the pytorch library\footnote{\url{http://pytorch.org/}} for our experiments. For training, we use Adam optimizer to minimize the weighted binary cross-entropy loss. We use weighted binary cross-entropy to partially mitigate the class imbalance issue mentioned previously. We use a weight of $0.2$ for negative samples and $0.8$ for positive samples. For Adam optimizer we use the most common setting with a learning rate of 0.001, $\beta_1 = 0.9$ and $\beta_2=0.999$. Training was performed with shuffled mini-batches of size 500 and a dropout of 0.2 was used for all layers. All the random initializations used Xavier normal distribution. We used $D = 100$(word embedding size), $C = 10$(context embedding size) and $M = 500$(number of topics). We used a single LSTM layer with 200 hidden states and Q = 100(classifier hidden layer size). We plan to make the source code publicly available.
 
\section{Results}

We evaluate our model on a held out set of 2801 documents. ROUGE metrics\cite{lin2004rouge} are used to compare the system generated extracts with the original abstracts of the papers. As explained previously summarizing scientific documents is a precision oriented task, and hence We report ROUGE-N precision (N=1,2,3,4). We compare our results with five widely accepted extractive summarization systems besides two state of art techniques. We specifically choose the topic signature\cite{lin2000automated} and latent semantic analysis\cite{steinberger2004using} based approaches due to their ability to identify the overall context and latent topics in a document. Besides these, we also compare our results with the popular graph based approaches, lexrank\cite{erkan2004lexrank} and textrank\cite{mihalcea2004textrank} and a simple frequency based approach. We also compare the results with Submodular optimization based technique\cite{lin2012learning} and Integer linear programming based summarization\cite{gillick2009scalable}, which are considered to be state of art techniques for sentence extraction\cite{hong2014repository}.\\

In order to make the results reproducible, we follow the guidelines suggested in \cite{hong2014repository} and use a fixed set of parameters when computing ROUGE scores\footnote{ROUGE-1.5.5 with the parameters: -n 4 -m -a -l 125 -x -c 95 -r 1000 -f A -p 0.5 -t 0}. Since the abstract size varies across documents in the evaluation set, we use the average abstract length of 125 words, when computing the ROUGE scores. All other parameters are same as those mentioned in \cite{hong2014repository}.  The results are shown in table \ref{table1} below.\\
\vspace{-5mm}
\begin{center}
	\begin{threeparttable}[H]
	\centering
	\caption{Results (ROUGE-N Precision)}
	\begin{tabular}{|c|c|c|c|c|}
		\hline
		Summarizer &      R-1       &      R-2       &           R-3            &           R-4            \\ \hline
		 Topicsum  &     0.266      &     0.055      &          0.020           &          0.012           \\
		   LSA     &     0.302      &     0.065      &          0.027           &          0.018           \\
		 LexRank   &     0.354      &     0.087      &          0.037           &          0.020           \\
		 TextRank  &     0.305      &     0.074      &          0.030           &          0.018           \\
		 FreqSum   &     0.331      &     0.088      &          0.034           &          0.018           \\
		Submodular & \textbf{0.360} &     0.087      &          0.036           &          0.022           \\
		   ILP     &     0.350           &  0.082              &   0.0355                       & 0.021 \\
		  Neural   &     0.344      & \textbf{0.090} & \textbf{0.042}$^\dagger$ & \textbf{0.027}$^\dagger$ \\ \hline
	\end{tabular}
	\begin{tablenotes}
	\small
	\item Figures in bold indicate the best performing system
	\item ${^\dagger}$ indicates significant difference with $\alpha=0.05$\\
	\end{tablenotes}
\label{table1}

\end{threeparttable}
\end{center}

As evident from table \ref{table1}, the proposed approaches outperforms all the existing systems on most ROUGE metrics. The only exception is Rouge-1 measure, where Submodular performs the best. We observe that R-3 and R-4 better reflect the systems ability to retain structural information in the abstract. A summary with good R-1 has more informative words but misses out on the structural information. A summary with higher R-3 or R-4 usually prefers sentences with clear cue phrases like 'results are significantly higher compared to' or 'in this paper we propose'. This is closer to the way a human would decide whether or not to include the information. A good R-1 does not necessitate that.

Below, we include a summary generated by our system along with the original abstract of the paper. The intensity of highlight shows the attention weights assigned by our model to a particular word. Darker the shade, higher the attention. In general, we observe that the proposed attention model efficiently identifies content words and cue phrase, both of which are important when selecting a sentence. For example, consider the first sentence of system generated summary: "In this paper we propose a statistical model for measure word generation for English-to-Chinese SMT systems, in which contextual knowledge from both source and target sentences is involved.". Our model identifies the phrases "In this paper we propose" (cue phrase) and " model for measure word" (part of title) as important. 
\vspace{-2mm}
\begin{InfoBox}[H]
\fbox{
\begin{minipage}{0.9\linewidth}		
\textbf{Document ID:} P08-1011
\vspace{1mm}

\textbf{Original Abstract: }		
Measure words in Chinese are used to indicate the count of nouns. Conventional statistical machine translation (SMT) systems do not perform well on measure word generation due to data sparseness and the potential long distance dependency between measure words and their corresponding head words. In this paper, we propose a statistical model to generate appropriate measure words of nouns for an English-to-Chinese SMT system. We model the probability of measure word generation by utilizing lexical and syntactic knowledge from both source and target sentences. Our model works as a post-processing procedure over output of statistical machine translation systems, and can work with any SMT system. Experimental results show our method can achieve high precision and recall in measure word generation.
\vspace{1mm}

\textbf{System generated summary: }
\noindent\fadingtext[left color = blue1, right color = blue2]{In this paper we propose a} statistical \fadingtext[left color = blue1, right color = blue2]{model for measure word} generation for English-to-Chinese SMT systems, in which \fadingtext[left color = blue1, right color = blue2]{ contextual} knowledge from both source and \fadingtext[left color = blue1, right color = blue2]{target sentences} is involved. \fadingtext[left color = blue1, right color = blue2]{To overcome the disadvantage of measure} word generation in a general SMT system, this paper proposes a dedicated \fadingtext[left color = blue1, right color = blue2]{statistical} model to generate measure words \fadingtext[left color = blue1, right color = blue2]{for English-to-Chinese} translation. \fadingtext[left color = blue1, right color = blue2]{Experimental results show} our method can significantly improve the quality of measure word generation. We also compared our method with a well known rule-based \fadingtext[left color = blue1, right color = blue2]{machine translation system - SYSTRAN3}. Most existing rule-based English-to-Chinese MT systems have a dedicated module handling measure word generation. 
\end{minipage}
}
\vspace{-2mm}
\caption*{Sample abstract and system generated summary}
\end{InfoBox}
\vspace{-4mm}
It is also interesting to note that the proposed model efficiently captures overall structure of the document, it starts with proposed work, then some details about experiment and system comparison. Barring the last sentence, it is quite precise and coherent in terms of content. Although it is not always possible to have sentences in the original documents that can directly be included in the abstract, results of the current experiment are quite encouraging and can serve as a very good first step towards abstract generation. 

\section{Conclusions}

In this work we proposed a weakly-supervised approach for generating extracts from scientific articles. We use topic models to create a context embedding that defines scope of the article and then use an attention based sequence encoder to generate sentence encoding. We then use pseudo labelled data to train a classifier that predicts whether or not a given sentence is \emph{summary worthy}. When evaluating on ACL anthology corpus, we were able to outperform the existing baseline and state of art techniques on ROUGE-2,3 and 4 metrics, while achieving a comparable performance on ROUGE-1. Moreover, we also demonstrate that our approach well preserves the overall structure of original document resulting in a final summary that is quite coherent. We envision this as a first step towards automatically creating abstracts of scientific articles and the results can be further used by generative techniques.

\bibliographystyle{ACM-Reference-Format}
\bibliography{sigir.bib}
\end{document}